\pdfoutput=1

\documentclass[reprint,aip,jcp,showpacs,twocolumn]{revtex4-1}
\usepackage{mathptmx}
\usepackage{epsfig,amsopn}
\usepackage{graphicx}
\usepackage{amsmath,amssymb}
\usepackage{natbib}
\usepackage{xcolor}
\usepackage{braket}
\usepackage{float}
\usepackage{enumerate}
\usepackage{mathtools}
\usepackage[symbol]{footmisc}
\usepackage{appendix}
\usepackage{stackengine}
\allowdisplaybreaks[1]

\newcommand{\eref}[1]{Eq.~(\ref{#1})}%
\newcommand{\fref}[1]{Fig.~\ref{#1}} %

\def\bea{\begin{eqnarray}}
\def\eea{\end{eqnarray}}

\def\p{\partial}

\begin{document}

\title{Space-dependent diffusion with stochastic resetting: A first-passage study}

\author{Somrita Ray}
\email{somritaray@mail.tau.ac.il}

\affiliation{\textit{
\noindent School of Chemistry, The Raymond and Beverly Sackler Center for Computational Molecular and Materials Science, The Center for Physics and Chemistry of Living Systems, \& The Ratner Center for Single Molecule Science, \\Tel Aviv University, Tel Aviv 69978, Israel}}

\date{\today}
\begin{abstract}
\noindent
We explore the effect of stochastic resetting on the first-passage properties of space-dependent diffusion in presence of a constant bias. In our analytically tractable model system, a particle diffusing in a linear potential $U(x)\propto\mu |x|$ with a spatially varying diffusion coefficient $D(x)=D_0|x|$ undergoes stochastic resetting, i.e., returns to its initial position $x_0$ at random intervals of time, with a constant rate $r$. Considering an absorbing boundary placed at $x_a<x_0$, we first derive an exact expression of the survival probability of the diffusing particle in the Laplace space and then explore its first-passage to the origin as a limiting case of that general result. In the limit $x_a\to0$, we derive an exact analytic expression for the first-passage time distribution of the underlying process. Once resetting is introduced, the system is observed to exhibit a series of dynamical transitions in terms of a sole parameter, $\nu\coloneqq(1+\mu D_0^{-1})$, that captures the interplay of the drift and the diffusion. Constructing a full phase diagram in terms of $\nu$, we show that for $\nu<0$, i.e., when the potential is strongly repulsive, the particle can never reach the origin. In contrast, for weakly repulsive or attractive potential ($\nu>0$), it eventually reaches the origin. Resetting accelerates such first-passage when $\nu<3$, but hinders its completion for $\nu>3$. A resetting transition is therefore observed at $\nu=3$, and we provide a comprehensive analysis of the same. The present study paves the way for an array of theoretical and experimental works that combine stochastic resetting with inhomogeneous diffusion in a conservative force-field.

\end{abstract}


\pacs{05.40.-a,05.40.Jc}


\maketitle
\section{Introduction}

Stochastic resetting implies a situation, where an ongoing dynamical process is stopped at random time epochs to start anew \cite{D1,D2,D4,ReviewSNM,ReuveniPRL,PalReuveniPRL}. It has gained overwhelming attention in recent times because of its spontaneous ubiquity in numerous natural and man made systems. For example, stock market crashes may reset the asset prices by drastically reducing those to some prior values \cite{economics}. Epidemics and natural disasters may have a similar effect on the population of a living species in a certain locality \cite{population1}. Search processes
may also reset \cite{FPUR1,FPUR2,FPUR5,FPUR7}; examples include foraging animals returning to their habitats \cite{forage,HomeRangeSearch1} due to fatigue or extreme weather. In computer science, it has long been known that resetting certain algorithms may significantly enhance their performance by reducing the effective run times \cite{CS1,CS2,CS3,CS4}. At the microscopic level, resetting is an indispensable part of the classical Michaelis–Menten reaction scheme \cite{Restart-Biophysics1, Restart-Biophysics2, Restart-Biophysics3, Restart-Biophysics4} and therefore, is crucial to the understanding of a variety of cellular processes\cite{Gunawardena,Rol,FD1}. For all these reasons and others, resetting and its applications have created a central point of scientific interest in recent years.
\\
\indent 
Diffusion with stochastic resetting serves as a classic model to explore resetting phenomena, where the completion of a first-passage process is accelerated due to resetting \cite{D1,D2,D4,D5,D7,D8,D12,D13,D15,RExp1,RExp2}. When the diffusion occurs in the presence of a bias, resetting either facilitates or hinders the resulting first-passage process \cite{ReuveniPRL,PalReuveniPRL}. As system parameters are varied, resetting may invert its role, which leads to a resetting transition \cite{Restart-Biophysics1,Restart-Biophysics2,exponent,Landau}. In recent years, diffusion with resetting in various potential landscapes have thoroughly been explored \cite{potential1,potential2,exponent,RayReuveniJPhysA,RayReuveniJCP}. In all these studies, however, the diffusion is assumed to be independent of the position of the particle.\\
\indent
Space-dependent or inhomogeneous diffusion \cite{vankampen,risken,lubensky,EliBarkai} frequently arises in a number of soft-matter systems. For instance, diffusion of tRNA inside the ribosome is found to be position-dependent in a recent study\cite{sdd6}. The diffusion coefficient of a Brownian particle in the vicinity of a wall or surface is greatly reduced due to hydrodynamic interactions \cite{sdd1,sdd2,sdd4,sdd5,sddw1,sddw3,softmatter1} and the mutual diffusion coefficient of two particles in a suspension depends on their separating distance \cite{sdd8}. A particle in geometric confinement undergoes diffusion that depends on its position, e.g., colloidal particles in porous media \cite{pm1,pm2,pm3}, particles trapped in vesicles \cite{sddv1} or in between two {\it nearly} parallel walls \cite{ostrowsky1,ostrowsky2,ostrowsky3}. Brownian particles confined in a narrow channel with uneven boundaries\cite{nc1,nc2,nc3,nc4,nc5,nc6} or inside a helical tube\cite{ht1,ht2} experience an effective space-dependent diffusivity along its direction of transport. Diffusion of a colloidal particle in a reversible chemical polymer gel \cite{sdd11}, micro-magnetic dynamics in ferromagnetic systems \cite{sdd12}, and reaction-diffusion inside a narrow channel\cite{nc7} also generate space-dependent diffusivity. Other popular examples of heterogeneous diffusion include dynamics of fluid membranes \cite{sdd9} and entangled polymer suspensions \cite{sdd10}. In a separate context, the presence of a space-dependent (multiplicative) noise term in Brownian dynamics has been found to manifest noise induced transitions \cite{sdd13,nt1,nt2,nt3}, asymmetric localization of particles\cite{nt4} and many other interesting transport phenomena \cite{t1,t3}.\\
\indent
The numerous examples of space-dependent diffusion in soft-matter systems made us curious to explore the effect of resetting on inhomogeneous diffusion process. Further motivation came from the recent experimental advances that successfully implemented a framework of resetting in a laboratory set up using optical tweezers to monitor a system of colloidal particles\cite{RExp1,RExp2}. Combined, these two fields, i.e, space-dependent diffusion and stochastic resetting, thus open up a horizon of possibilities. As the present work is the first step in that direction, here our aim is to provide an in-depth analysis of the possible effects of Poissonian resetting on an exactly solvable model system, where a particle in a linear potential $U(x)\propto \mu|x|$ diffuses inhomogeneously in space with a diffusion coefficient $D(x)=D_0|x|$. The space-dependent nature of the diffusion reduces the fluctuations of motion as the particle approaches the origin, while such fluctuations are enhanced when it moves away from the origin. To explore the possible effects that resetting may have on the first-passage properties of the present system, we consider
a single governing parameter $\nu\coloneqq (1+\mu D_0^{-1})$ that captures the interplay of the drift and diffusion. Assuming an absorbing boundary at the origin, we show that the particle can reach that boundary only when $\nu>0$ and resetting accelerates the resulting first-passage for $0<\nu<3$. In complete contrast, for $\nu\ge 3$, the introduction of resetting delays the mean completion time of the process. Summarizing these results we construct a phase diagram [see \fref{Fig5}], where transitions between the different dynamical behaviors are observed by tuning $\nu$.\\
\indent
The rest of this paper is organized as follows. We start in Sec. II where we consider a particle undergoing space-dependent diffusion with resetting in presence of a constant bias and study its first-passage to an absorbing boundary. In particular, we derive an exact expression of the survival probability of the particle in the Laplace space and utilize the same to calculate the first-passage time. The results obtained in Sec. II hold for any arbitrary position of the absorbing boundary, provided it is placed between the origin and the initial position of the particle. In Sec. III, we explore the limiting case, where the absorbing boundary is placed at the origin. There, we first study the underlying process and derive an exact analytical expression for the first-passage time distribution. In the same Section, we investigate the effect of resetting on the system when it diffuses to the origin. The resetting transition is discussed in Sec. IV. In Sec. V, we construct a full phase diagram for the present problem and draw the final conclusions in Sec. VI.
\section{Space-dependent diffusion: First-passage with resetting}
\subsection{The model}
\begin{figure}[t!]
\begin{centering}
\includegraphics[width=8.30cm]{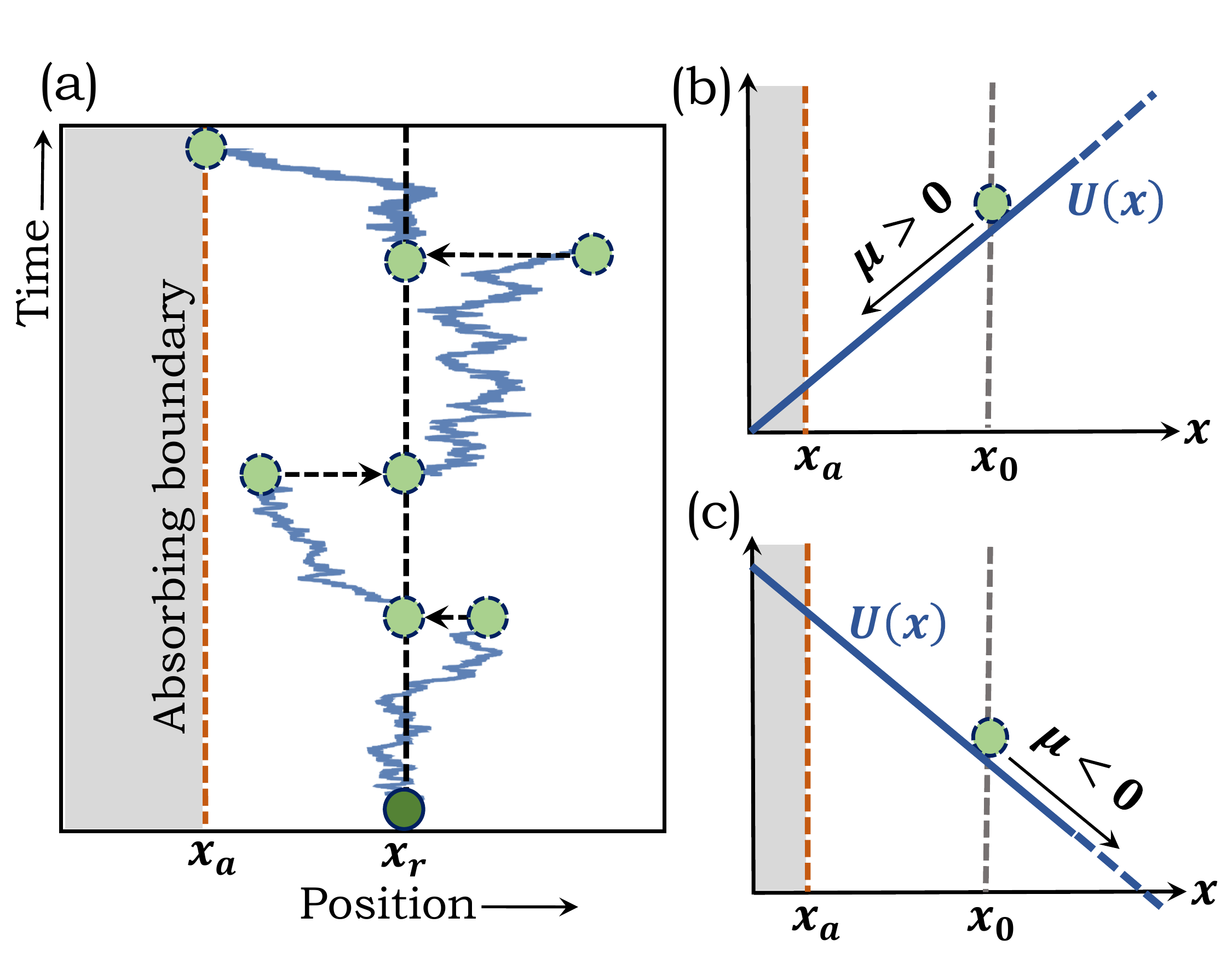}
\end{centering}
\caption{An illustrative model of the system, where a particle in a potential $U(x)\propto \mu|x|$ undergoes space-dependent diffusion in the interval $x\in[0,\infty)$ with a diffusion coefficient $D(x)=D_0|x|$.  Panel (a): A schematic trajectory of the particle that experiences stochastic resetting at $x_r$ while diffusing toward an absorbing boundary at $x_a\ge 0$. Panels (b): The drift velocity $\mu$ experienced by the particle in an attractive linear potential. Panel (c): The drift velocity $\mu$ experienced by the particle in a repulsive linear potential.}
\label{Fig1}
\end{figure}

Consider a particle diffusing in a linear potential $U(x)=U_0|x|$ with a space-dependent diffusion coefficient that varies linearly with the distance from the origin as $D(x)=D_0|x|$, where $D_0>0$ is the proportionality constant. These special choices of $D(x)$ and $U(x)$ ensure that the particle, with an initial position $x_0>0$, can never cross the origin. Therefore, the system is essentially confined at the positive values of $x$. Note that when $U_0>0$, the potential is attractive, whereas it is repulsive for $U_0<0$. Assume that the particle is stochastically reset to a position $x_r>0$ [see \fref{Fig1}(a)] with a constant rate $r$, which implies that the random times between two consecutive resetting events are drawn from an exponential distribution with mean $1/r$. Letting $p_r(x,t|x_0)$ denote the conditional probability density of finding the particle at position $x$ at time $t$, provided that the initial position was $x_0$, the Fokker-Planck equation for the process with resetting \cite{ReviewSNM} can be written as
\begin{eqnarray}
\dfrac{\partial p_r(x,t|x_0)}{\partial t}=
\mu\dfrac{\partial }{\partial x} p_r(x,t|x_0)
+D_0\dfrac{\partial^2}{\partial x^{2}}\left[x p_r(x,t|x_0)\right]\nonumber \\
-r p_r(x,t|x_0)+r\delta(x-x_r),
\label{eq:fper}
\end{eqnarray}

\noindent
where $\delta(x-x_r)$ is a Dirac delta function. Here $\mu\coloneqq U_0\zeta^{-1}$ is the constant drift velocity, $\zeta$ being the friction coefficient. Since $\zeta$ is always positive, the drift acts towards the origin ($\mu>0$) when the potential is attractive and away from the origin ($\mu<0$) when the potential is repulsive [see \fref{Fig1}(b),(c)]. Note that when $r=0$, \eref{eq:fper} boils down to the Fokker-Planck equation for the underlying process without resetting. Once resetting is introduced, i.e., for $r>0$, there is a loss of probability from position $x$ and a subsequent gain of probability at position $x_r$. The last two terms on the right hand side of \eref{eq:fper} account for this additional probability flow, which is proportional to $r$, the rate of resetting. \\
\indent
Consider an absorbing boundary at $x_a<x_0$ [see \fref{Fig1}], which implies that when the particle, starting at $x_0>0$, hits that boundary for the first time, it is immediately removed from the system, leading to $p_r(x_a,t|x_0)=0$. In terms of the survival probability $Q_r(t|x_0)\coloneqq\int_{\Omega} p_r(x,t|x_0) dx$, i.e, the probability that the particle exists in the interval $\Omega=[x_a,\infty)$ at time $t$, the backward Fokker Planck equation\cite{gardiner,FPT1,FPT2,ReviewSNM} for the above process is given by
\begin{align}
\dfrac{\partial Q_r(t|x_0)}{\partial t}=
-\mu\dfrac{\partial  Q_r(t|x_0)}{\partial x_0}
+D_0 x_0\dfrac{\partial^2 Q_r(t|x_0)}{\partial x_0^{2}}
\nonumber\\
-r Q_r(t|x_0)+rQ_r(t|x_r),
\label{eq:bfpeQr}
\end{align}
with the initial condition $Q_r(0|x_0) = 1$ and the boundary condition $Q_r(t|x_a) = 0$. Laplace transforming \eref{eq:bfpeQr} we obtain
\begin{align}
D_0 x_0\dfrac{\partial^2 \tilde{Q_r}(s|x_0)}{\partial x_0^{2}}-\mu\dfrac{\partial  \tilde{Q_r}(s|x_0)}{\partial x_0}- (s+r) \tilde{Q_r}(s|x_0)=\nonumber\\-\left[1+r\tilde{Q_r}(s|x_r)\right],
\label{eq:Qr_lt}
\end{align}
where $\tilde{Q_r}(s|x_0)\coloneqq \int_0^{\infty}e^{-st}Q_r(t|x_0)dt$ denotes the Laplace transform of $Q_r(t|x_0)$. Letting $T_r$ denote the first-passage time (FPT) to the absorbing boundary placed at $x_a$, we recall that the probability density of $T_r$ is given by $-\partial Q_r(t|x_0)/\partial t$\cite{gardiner,FPT1}. This allows us to calculate any moment of $T_r$ from $\tilde{Q_r}(s|x_0)$ following the relation
\begin{eqnarray}
\left<T_r^n\right>&=&\int_{0}^{\infty}t^n\left[-\frac{\p Q_r(t|x_0)}{\p t}\right]dt\nonumber\\
&\equiv&n\left(-1\right)^{n-1}\left[\frac{d^{n-1}\tilde{Q_r}(s|x_0)}{ds^{n-1}}\right]_{s\to0}.
\label{eq:fpt_moments}
\end{eqnarray}
Since we are interested in the first-passage properties of the system, next we solve \eref{eq:Qr_lt} in order to find out the survival probability in the Laplace space.
\vspace{-0.2cm}
\subsection{The survival probability}
\eref{eq:Qr_lt} is a linear non-homogeneous differential equation. In order to convert it to a homogeneous one, we first consider a constant shift as 
\begin{equation}
\tilde{q}_r(s|x_0)\coloneqq\tilde{Q}_r(s|x_0)-\left[\frac{1+r\tilde{Q}_r(s|x_r)}{s+r}\right].
\label{eq:qr_def}
\end{equation}
\eref{eq:Qr_lt} in terms of $\tilde{q}_r(s|x_0)$ reads
\begin{align}
D_0 x_0\dfrac{\partial^2 \tilde{q_r}(s|x_0)}{\partial x_0^{2}}-\mu\dfrac{\partial  \tilde{q_r}(s|x_0)}{\partial x_0}- (s+r) \tilde{q_r}(s|x_0)=0.
\label{eq:qr_lt}
\end{align}
Performing a variable transformation as $\rho(x_0)\equiv\sqrt{x_0}$ and considering that $\tilde{q_r}(s|\rho)\equiv \rho^{\nu} \tilde{y}(s|\rho)$, where $\nu$ is an arbitrary constant, we can rewrite \eref{eq:qr_lt} in terms of $\tilde{y}(s|\rho(x_0))$ as 
\begin{eqnarray}
\dfrac{\partial^2 \tilde{y}(s|\rho)}{\partial \rho^{2}}+\left(\frac{c_1}{\rho}\right)\dfrac{\partial  \tilde{y}(s|\rho)}{\partial\rho}+\left[\frac{c_2}{\rho^2}-4\left(\frac{s+r}{D_0}\right)\right]\tilde{y}(s|\rho)=0,\nonumber\\
\label{eq:y_rho}
\end{eqnarray}
where $c_1=2\nu-1-2\mu D_0^{-1}$ and $c_2=\nu[\nu-2(1+\mu D_0^{-1})]$. Assigning $c_1=1$, we get $\nu=(1+\mu D_0^{-1})$, which in turn leads to $c_2=-\nu^2$. Therefore, \eref{eq:y_rho} reduces to
\begin{eqnarray}
\dfrac{\partial^2 \tilde{y}(s|\rho)}{\partial \rho^{2}}+\left(\frac{1}{\rho}\right)\dfrac{\partial  \tilde{y}(s|\rho)}{\partial \rho}=\left[\left(\frac{\nu}{\rho}\right)^2+4\left(\frac{s+r}{D_0}\right)\right] \tilde{y}(s|\rho).\nonumber\\
\label{eq:y_rho2}
\end{eqnarray}
\eref{eq:y_rho2} is a modified Bessel equation with general solution\cite{BesselSol} 
\begin{align}
\tilde{y}(s|\rho)=
\begin{cases}
A_+I_{\nu}\left(2\sqrt{\frac{s+r}{D_0}}\rho\right)+B _+K_{\nu}\left(2\sqrt{\frac{s+r}{D_0}}\rho\right)\;\;\mbox{if}\;\nu>0 \\ \\
A_-I_{-\nu}\left(2\sqrt{\frac{s+r}{D_0}}\rho\right)+B_- K_{-\nu}\left(2\sqrt{\frac{s+r}{D_0}}\rho\right)\;\;\mbox{if}\;\nu<0.
\end{cases}
\label{eq:y_sol}
\end{align}
Here $I_{\nu}(y)\coloneqq\sum_{k=0}^{\infty}\frac{1}{\Gamma(k+\nu+1)k!}\left(\frac{y}{2}\right)^{2k+\nu}$ is the modified Bessel function of the first kind \cite{NIST} and $K_{\nu}(y)=\frac{\pi}{2}\left[\frac{I_{-\nu}(y)-I_{\nu}(y)}{\sin{(\nu\pi)}}\right]$ is the modified Bessel function of the second kind \cite{NIST}, defined in terms of $I_{\nu}(\cdot)$. \\
\indent
Recalling that $\rho(x_0)=\sqrt{x_0}$ and $\tilde{q_r}(s|x_0)=x_0^{\nu/2} \tilde{y}(s|x_0)$, from \eref{eq:qr_def} and \eref{eq:y_sol} we obtain the general solution of \eref{eq:Qr_lt} as
\begin{widetext}
\begin{eqnarray}
\tilde{Q_r}(s|x_0)=\begin{cases}
A_+x_0^{\frac{1}{2}\left(1+\frac{\mu}{D_0}\right)}
I_{1+\frac{\mu}{D_0}}\left(2\sqrt{\frac{s+r}{D_0}} \sqrt{x_0}\right)+
B_+x_0^{\frac{1}{2}\left(1+\frac{\mu}{D_0}\right)}
K_{1+\frac{\mu}{D_0}}\left(2\sqrt{\frac{s+r}{D_0}} \sqrt{x_0}\right)
+\left[\frac{1+r\tilde{Q_r}(s|x_r)}{s+r}\right]\;\;\;\;\;\;\mbox{if}\;\left(1+\frac{\mu}{D_0}\right)>0,
\\ \\
A_-x_0^{\frac{1}{2}\left(1+\frac{\mu}{D_0}\right)}
I_{-1-\frac{\mu}{D_0}}\left(2\sqrt{\frac{s+r}{D_0}} \sqrt{x_0}\right)+
B_-x_0^{\frac{1}{2}\left(1+\frac{\mu}{D_0}\right)}
K_{-1-\frac{\mu}{D_0}}\left(2\sqrt{\frac{s+r}{D_0}} \sqrt{x_0}\right)
+\left[\frac{1+r\tilde{Q_r}(s|x_r)}{s+r}\right]\;\;\;\;\;\;\mbox{if}\;\left(1+\frac{\mu}{D_0}\right)\le0.
\end{cases}
\label{eq:Qr_sol}
\end{eqnarray}
\end{widetext}
\noindent 
In order to obtain the specific solution of \eref{eq:Qr_lt} from \eref{eq:Qr_sol}, we need to find out the explicit expressions of $A_{\pm}$ and $B_{\pm}$ from the boundary conditions. \\
\indent
Since $\tilde{Q_r}(s|x_0)$ should be finite even at $x_0\to \infty$,  we set $A_{\pm}=0$. The absorbing boundary at $x_a$ leads to $\tilde{Q_r}(s|x_a)=0$, which gives
\begin{align}
B_{\pm}=-\left[\frac{1+r\tilde{Q_r}(s|x_r)}{s+r}\right]\frac{x_a^{\frac{1}{2}\left(1+\frac{\mu}{D_0}\right)}}{K_{\pm\left(1+\frac{\mu}{D_0}\right)}\left(2\sqrt{\frac{s+r}{D_0}} \sqrt{x_0}\right)}.
\label{eq:Qr_abs}
\end{align}
Note that $K_{1+\frac{\mu}{D_0}}(\cdot)=K_{-1-\frac{\mu}{D_0}}(\cdot)$, which leads to $B_+=B_-$. Substituting $A\equiv A_{\pm}$ and $B\equiv B_{\pm}$ in \eref{eq:Qr_sol} we get
\begin{align}
\tilde{Q_r}(s|x_0)=\frac{1+r\tilde{Q_r}(s|x_r)}{s+r}
\left[1-\frac{x_0^{\frac{\nu}{2}}K_{\nu}\left(2\sqrt{\frac{s+r}{D_0}} \sqrt{x_0}\right)}{x_a^{\frac{\nu}{2}}K_{\nu}\left(2\sqrt{\frac{s+r}{D_0}} \sqrt{x_a}\right)}\right],
\label{eq:Qr_self}
\end{align}
where $\nu\coloneqq (1+\mu D_0^{-1})$, as obtained earlier. Note that for attractive potential $\nu>1$, whereas for repulsive potential $\nu<1$. Considering that the particle is reset to its initial position, we set $x_r=x_0$ in \eref{eq:Qr_self} to obtain an explicit expression of $\tilde{Q_r}(s|x_0)$ in a self-consistent manner as
\begin{eqnarray}
\tilde{Q_r}(s|x_0)=\frac{1-\left[\frac{x_0}{x_a}\right]^{\frac{\nu}{2}}\frac{K_{\nu}\left(2\sqrt{\frac{s+r}{D_0}} \sqrt{x_0}\right)}{K_{\nu}\left(2\sqrt{\frac{s+r}{D_0}} \sqrt{x_a}\right)}}
{s+r
\left[\left(\frac{x_0}{x_a}\right)^{\frac{\nu}{2}}\frac{K_{\nu}\left(2\sqrt{\frac{s+r}{D_0}} \sqrt{x_0}\right)}{K_{\nu}\left(2\sqrt{\frac{s+r}{D_0}} \sqrt{x_a}\right)}\right]}.
\label{eq:Qr_1}
\end{eqnarray}
\eref{eq:Qr_1} thus presents an exact expression of the survival probability (in the Laplace space) of the particle in the interval $\Omega=[x_a,\infty)$ while it diffuses in a linear potential $U(x)\propto \mu|x|$ with a diffusion coefficient $D(x)=D_0|x|$ in presence of resetting. In what follows, we will proceed to explore $T_r$, the first-passage time to the absorbing boundary placed at $x_a$, with the aid of \eref{eq:Qr_1}.
\subsection{The first-passage time: Mean and standard deviation}
Recalling \eref{eq:fpt_moments}, we see that the mean FPT from $x_0$ to $x_a$ in presence of resetting can be obtained as $\left<T_r\right>=\left[\tilde{Q_r}(s|x_0)\right]_{s=0}$. Setting $s=0$ in \eref{eq:Qr_1} leads to
\begin{align}
\left<T_r\right>=\frac{1}{r}\left[\left(\frac{x_a}{x_0}\right)^{\frac{\nu}{2}}\left[\frac{K_{\nu}(2\sqrt{\frac{r}{D_0}}\sqrt{x_a})}{K_{\nu}(2\sqrt{\frac{r}{D_0}}\sqrt{x_0})}\right]-1\right].
\label{eq:MFPT_1}
\end{align}\\
\noindent
In a similar spirit, the second moment of $T_r$ is obtained following \eref{eq:fpt_moments} as $\left<T_r^2\right>=-2 [\partial \tilde{Q}_r(s|x_0)/\partial s]_{s=0}$. Utilizing that relation and setting $\alpha\coloneqq 2\sqrt{r/D_0}$, we calculate the standard deviation of the FPT, $\sigma(T_r)\coloneqq \sqrt{\left<T_r^2\right>-\left<T_r\right>^2}$, that reads   

\begin{widetext}
\begin{eqnarray}
\sigma(T_r)=
\frac{1}{r}\sqrt{
\left[\frac{\alpha\sqrt{x_a}
K_{\nu-1}\left(\alpha \sqrt{x_a}\right)
K_{\nu}\left(\alpha \sqrt{x_0}\right)+
K_{\nu}\left(\alpha \sqrt{x_a}\right)
\left[\left(\frac{x_a}{x_0}\right)^{\frac{\nu}{2}}K_{\nu}\left(\alpha \sqrt{x_a}\right)
-\alpha\sqrt{x_0}
K_{\nu-1}\left(\alpha \sqrt{x_0}\right)
\right]}
{\left(\frac{x_0}{x_a}\right)^{\frac{\nu}{2}}\left[K_{\nu}\left(\alpha \sqrt{x_0}\right)\right]^2}\right]-1}.
\label{eq:STDV_1}
\end{eqnarray}
\end{widetext}
\vspace{0.2cm}

\begin{figure}[t!]
\begin{centering}
\includegraphics[width=7.8cm]{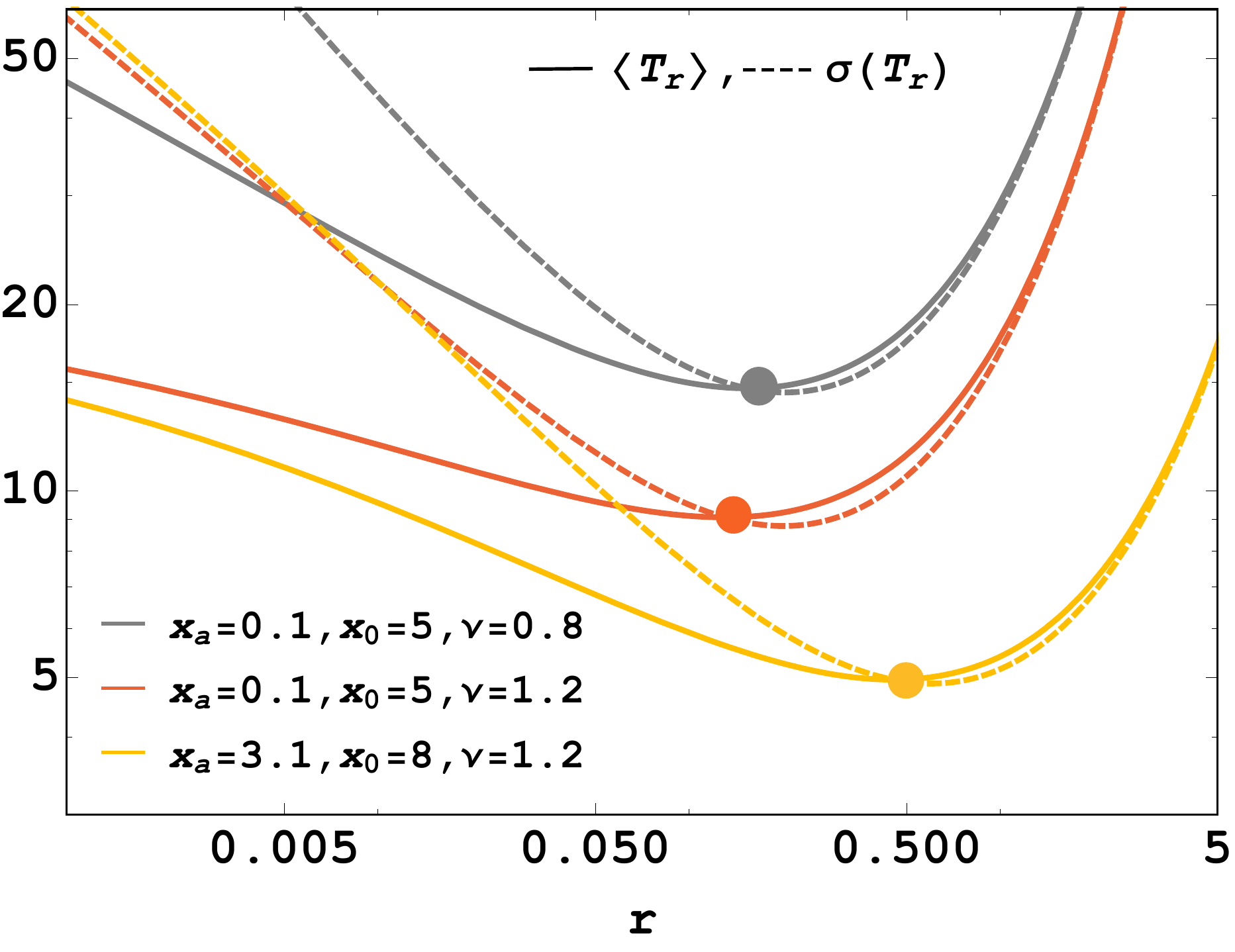}
\end{centering}
\caption{The mean $\left<T_r\right>$ from \eref{eq:MFPT_1} and the standard deviation $\sigma(T_r)$ from \eref{eq:STDV_1} with the resetting rate $r$ for different values of the system parameter $\nu$, $x_a$ and $x_0$. The colored circles mark the minimum value of $\left<T_r\right>$ for each parameter set. Here we have taken $D_0=1$. The distance between the initial position $x_0$ and the position of the absorbing boundary $x_a$ is kept conserved as $(x_0-x_a) = 4.9$.}
\label{Fig2}
\end{figure}

\indent
The results derived in this Section hold for space-dependent diffusion with resetting to an absorbing boundary placed at any arbitrary position $x_a<x_0$ in presence of a constant bias. In \fref{Fig2} we plot the mean FPT from \eref{eq:MFPT_1} and the standard deviation from \eref{eq:STDV_1} with the resetting rate $r$ for different values of $\nu$, $x_a$ and $x_0$, where we keep the distance between $x_0$ and $x_a$ constant. It appears from \fref{Fig2} that for our choice of parameters, $\left<T_r\right>$ exhibits a non-monotonic variation with $r$. 
This implies that by resetting the particle at its initial position $x_0$ with a suitable rate, it is possible to significantly lower the mean FPT to an absorbing boundary placed at $x_a<x_0$. For highly frequent resetting events, the particle fails to reach the boundary in finite mean time, which explains the divergence of $\left<T_r\right>$ for higher values of $r$. In an analogous way, the non-monotonic variation in $\sigma(T_r)$ signifies that the fluctuations in FPT can also be considerably reduced by resetting the system. In addition, \fref{Fig2} indicates that such acceleration of first-passage due to resetting can, in principle, be observed when the potential is either repulsive ($\nu<1$) or attractive ($\nu>1$).\\
\indent
\fref{Fig2} suggests that the effect of resetting should depend significantly on the placement of the absorbing boundary with respect to the origin, a special feature that can be attributed to the inhomogeneous nature of the diffusion. Recalling that the  diffusion coefficient $D(x)$ varies linearly with $x$, we see that it vanishes at the origin. Hence, as the particle moves close to the origin, its dynamics gets drift-dominated. Therefore, it is not expected to ever reach the origin when the potential is repulsive. Introduction of resetting to the system might help the particle reach the origin in this case. In stark contrast, when the potential is attractive, the particle should always reach the origin; resetting can either accelerate or delay such first-passage. Note that once the particle reaches the origin, the attractive potential will not allow it to leave, hence in absence of the absorbing boundary it is expected to stay there forever. Motivated by the above possibilities, in the rest of this paper we perform a comprehensive analysis of the first-passage of the system to the origin.  
\section{Reaching the origin}
In this Section, we explore the first-passage of the particle to the origin as a limiting case ($x_a\to 0$) of the general results obtained in the previous Section. To start with, we consider the survival probability from \eref{eq:Qr_1} in the limit $x_a\to 0$. The limiting expression of the modified Bessel function $K_{\nu}(\cdot)$ for small arguments\cite{NIST} for $\nu\le0$ gives $\lim_{x_a\to 0}x_a^{-\nu/2}/K_{\nu}(a\sqrt{x_a})= 0$, which leads to $\tilde{Q}_r(s|x_0)= s^{-1}$, i.e., $Q_r(t|x_0)=1$. Therefore, the survival probability is always conserved to unity for $\nu\coloneqq(1+\mu D_0^{-1})<0$, which means that the particle can never reach the origin when the linear potential is strongly repulsive ($\mu<-D_0$), even when it is reset at $x_0$ with a rate $r>0$.
Resetting is expected to lead to a non-equilibrium steady state in this case; we will address that elsewhere. \\
\indent
As we are interested in the first-passage properties of the system in the present work, our discussion will henceforth be restricted to $\nu\ge 0$.
In what follows, we start with the underlying process ($r\to 0$) to study its first-passage to the origin and then, introduce resetting to explore its effect on such first-passage.

\subsection{First-passage without resetting}
In the absence of resetting, i.e, for $r\to 0$, \eref{eq:Qr_1} boils down to 
\begin{eqnarray}
\tilde{Q_0}(s|x_0)&\coloneqq&\lim_{r\to 0}\tilde{Q_r}(s|x_0)\nonumber\\
&=&\frac{1}{s}
\left[1-\left[\frac{x_0}{x_a}\right]^{\frac{\nu}{2}}\frac{K_{\nu}\left(2\sqrt{\frac{s}{D_0}} \sqrt{x_0}\right)}{K_{\nu}\left(2\sqrt{\frac{s}{D_0}} \sqrt{x_a}\right)}\right],
\label{eq:Qr_r0}
\end{eqnarray}
where $\tilde{Q}_0(s|x_0)\coloneqq \int_0^{\infty}e^{-st}Q_0(t|x_0)dt$ is the survival probability of the underlying process in the Laplace space. For $\nu>0$, the limiting expression for the modified Bessel function $K_{\nu}(\cdot)$ for small arguments\cite{NIST} leads to $\lim_{x_a\to 0}x_a^{\nu/2}K_{\nu}(a\sqrt{x_a})\simeq 2^{\nu-1}\Gamma(\nu)/a^{\nu}$, where $\Gamma(\nu)\coloneqq \int_0^{\infty}t^{\nu-1}e^{-t}dt$ is the Gamma function. Plugging in this expression into \eref{eq:Qr_1}, we obtain
the survival probability in the limit $x_a\to 0$, which reads
\begin{align}
\tilde{Q}_0(s|x_0)=
\frac{1}{s}
\left[1-\frac{2
}{\Gamma(\nu)}\left[\frac{s x_0}{D_0}\right]^{\frac{\nu}{2}}K_{\nu}\left(2\sqrt{\frac{s x_0}{D_0}}\right)\right].
\label{eq:Q0r_r0}
\end{align}
Letting $T_0$ denote the FPT of the underlying process and recalling that the probability density of $T_0$ is given by\cite{gardiner,FPT1} $f_{T_0}(t)=-dQ_0(t|x_0)/dt$, we see that $\tilde{Q}_0(s|x_0)=[1-\tilde{f}_{T_0}(s)]/s$. Here $\tilde{f}_{T_0}(s)$ is the Laplace transform of $f_{T_0}(t)$. \eref{eq:Q0r_r0} thus gives
\begin{align}
\tilde{f}_{T_0}(s)\coloneqq
\int_0^{\infty}e^{-st}f_{T_0}(t)dt=
\frac{2
}{\Gamma(\nu)}\left[\frac{s x_0}{D_0}\right]^{\frac{\nu}{2}}K_{\nu}\left(2\sqrt{\frac{s x_0}{D_0}}\right).
\label{eq:T0_lt}
\end{align}
Next, we consider the following identity \cite{zwillinger}
\begin{align}
\int_0^{\infty}p^{-(\gamma+1)} \exp{\left[-p-\frac{\lambda^2}{4p}\right]}dp = 2\left(\frac{\lambda}{2}\right)^{\gamma}K_{\gamma}(\lambda),
\label{eq:zwilinger}
\end{align}
which holds for $|arg (\lambda)|<\pi/2$ and $Re(\lambda^2)>0$. Comparing Eqs.~(\ref{eq:T0_lt}) and (\ref{eq:zwilinger}), we identify $\gamma \equiv \nu$, $\lambda \equiv 2\sqrt{sx_0/D_0}$ and $p\equiv st$, which allows us to write 
\begin{equation}
f_{T_0}(t)=\frac{t^{-\left(\nu+1\right)}}{\Gamma(\nu)}\left(\frac{x_0}{D_0}\right)^{\nu}\exp{\left[-\frac{x_0}{D_0 t}\right]}.
\label{eq:fT0}
\end{equation} 
\eref{eq:fT0} thus presents the first-passage time distribution to the origin for a particle that undergoes space-dependent diffusion in a weakly repulsive or attractive linear potential ($\nu>0$). Note that in the long time limit $f_{T_0}(t)\sim t^{-(\nu+1)}$. Since $f_{T_0}(t)=-dQ_0(t|x_0)/dt$, the survival probability in the limit $t\to\infty$ decays as $Q_0(t|x_0)\sim t^{-\nu}$. Thus $\nu$ governs the decay of the survival probability of the underlying process, and hence can be identified as the ``persistence exponent"\cite{FPT2} for the present problem. It is evident from \eref{eq:fT0} the tail of the distribution $f_{T_0}$ gets heavier as $\nu$ decreases. Resetting the system with a suitable rate can then effectively shorten that heavy tail of the FPT distribution, thereby accelerating the resulting first-passage.
\begin{figure*}[ht!]
\begin{centering}
\includegraphics[width=4.5cm]{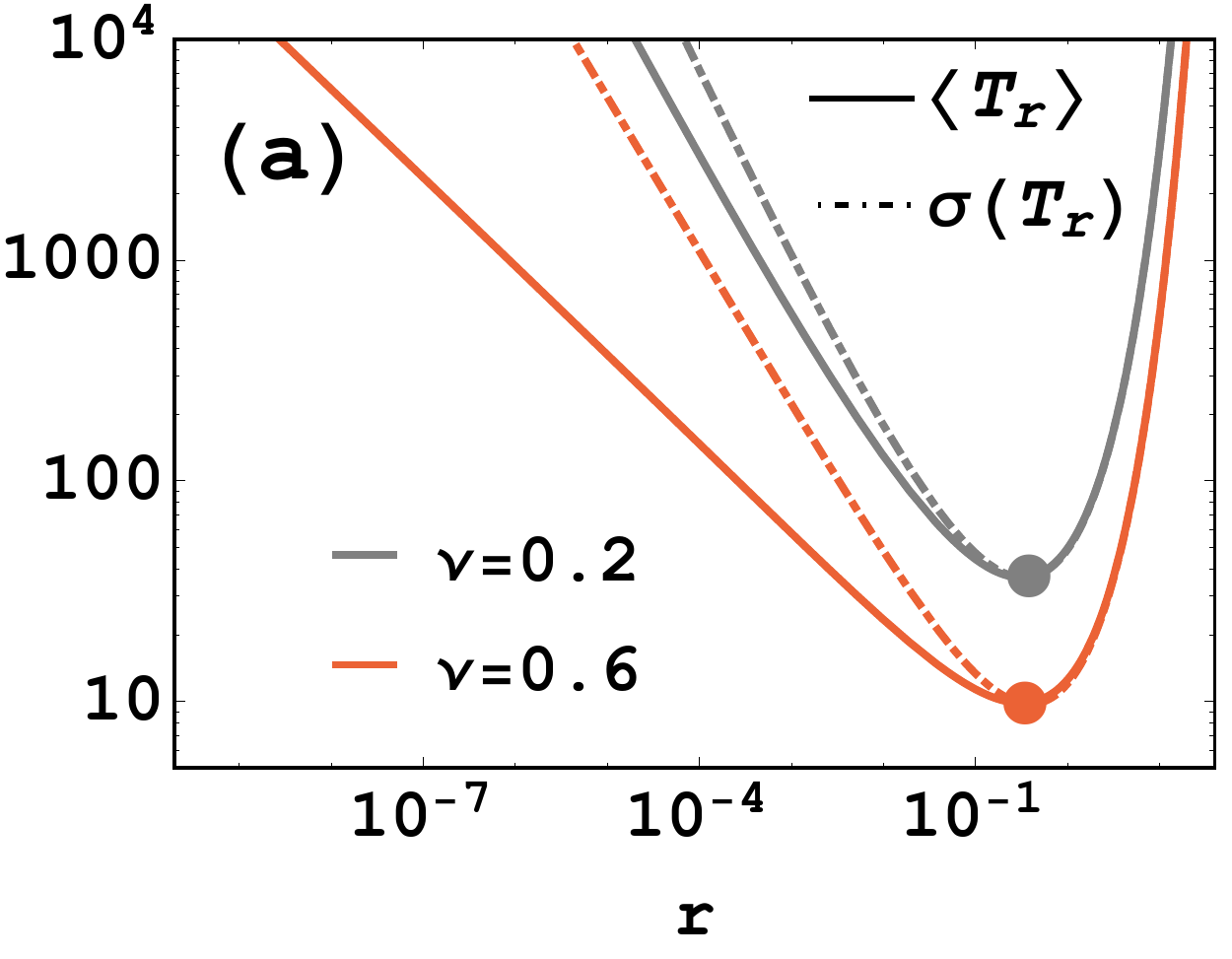}
\includegraphics[width=4.35cm]{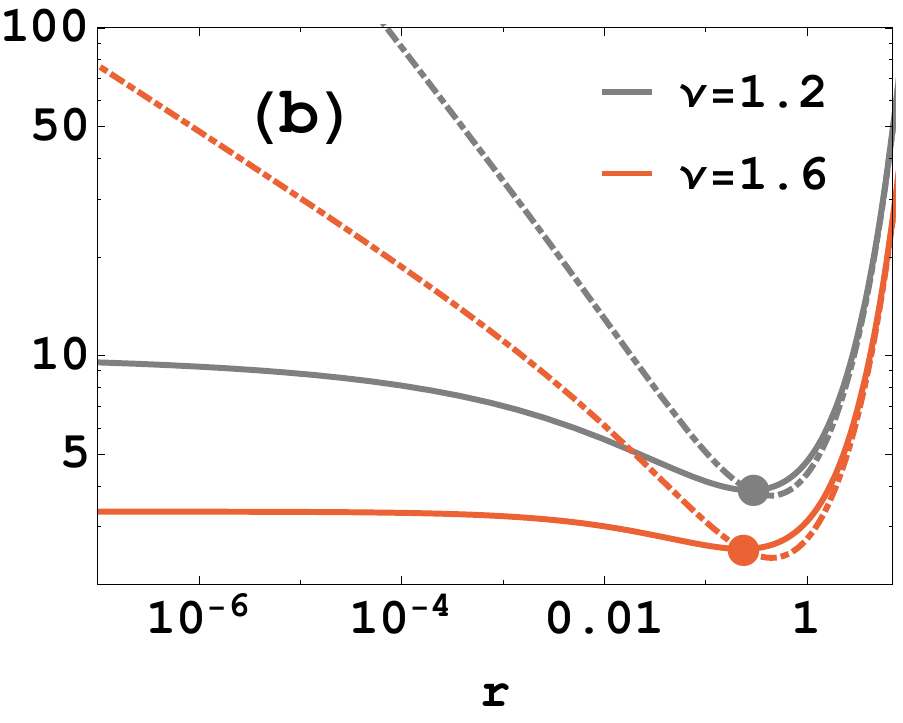}
\includegraphics[width=4.3cm]{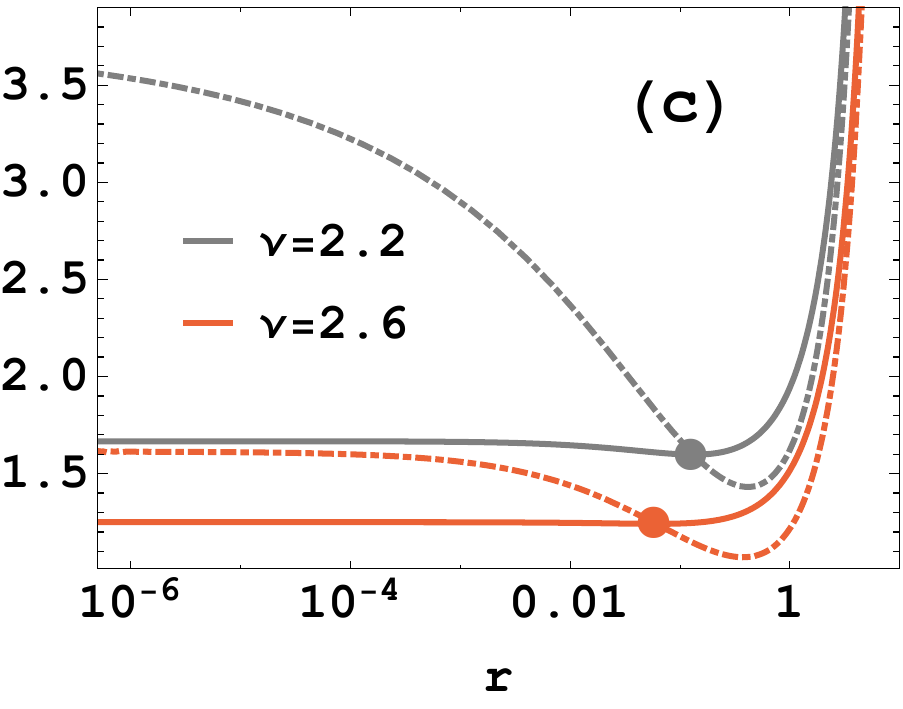}
\includegraphics[width=4.32cm]{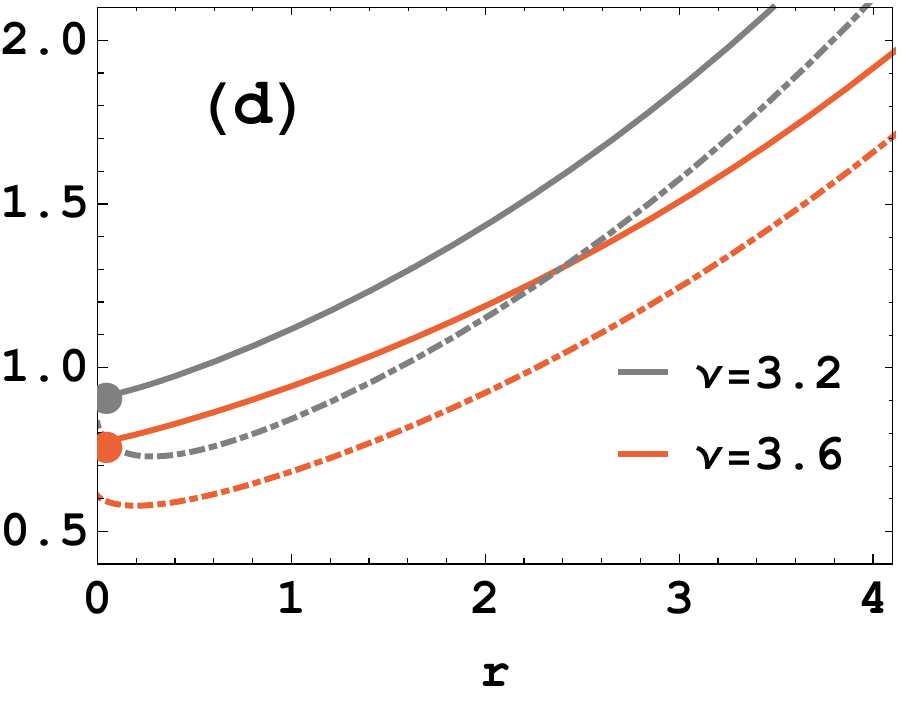}%
\end{centering}
\caption{The mean FPT $\left<T_r\right>$ [from \eref{eq:MFPT0}] and the standard deviation of the FPT $\sigma(T_r)$ [from \eref{eq:STDVr0}] vs. the resetting rate $r$, for four different phases of space-dependent diffusion in a linear potential. Solid lines indicate $\left<T_r\right>$ and dashed lines show $\sigma(T_r)$. Curves of similar color denote the same value of $\nu$, while the colored circles mark the minimum value of $\left<T_r\right>$ for each choice of $\nu$. 
Panel (a): For $0<\nu<1$, the mean $\left<T_r\right>$ and the standard deviation $\sigma(T_r)$ of the FPT, both diverge in the limit $r\to 0$ (no resetting). 
Panel (b): For $1<\nu<2$, while $\left<T_r\right>$ is finite in the limit $r\to 0$, $\sigma(T_r)$ diverges. 
Panel (c): For $2<\nu<3$, the mean and the standard deviation of FPT, both are finite for $r\to 0$, and  $\sigma(T_{r\to0})>\left<T_{r\to0}\right>$. 
Panel (d): For $\nu>3$, both the mean and standard deviation are finite in the limit $r\to 0$, and $\sigma(T_{r\to0})<\left<T_{r\to0}\right>$. Panels (a)$-$(c) show that for $0<\nu<3$, $\left<T_r\right>$ exhibits non-monotonic variation with the resetting rate, whereas panel (d) shows that for $\nu>3$, the mean FPT monotonically increases with $r$. In all panels we have taken $x_0 = 2.0$ and $D_0=1.0$.
}
\label{Fig3}
\end{figure*}

\indent
\eref{eq:fT0} indicates that the $n$-th moment of $T_0$, $\left<T_{0}^n\right>\coloneqq\int_0^{\infty}t^nf_{T_0}(t)dt$, diverges for $\nu\le n$, but is finite for $\nu>n$ and can be given by the general expression
\begin{align}
\left<T_{0}^n\right>=\left(\frac{x_0}{D_0}\right)^n\frac{\Gamma(\nu-n)}{ \Gamma(\nu)}\;\;\;\;\mbox{where}\;\;\;n=1,2,3,...
\label{eq:MFPTn_r0}
\end{align}
Note that the moments of $T_{0}$ can as well be calculated from \eref{eq:Q0r_r0} following the relations given in \eref{eq:fpt_moments} for the underlying process in the absence of resetting.\\
\indent
Setting $n=1$ in \eref{eq:MFPTn_r0}, the mean FPT to the origin in the absence of resetting can be obtained for $\nu>1$ as 
\begin{align}
\left<T_{0}\right>=\frac{x_0}{D_0 (\nu-1)}.
\label{eq:MFPT0_r0}
\end{align}
\noindent
In a similar spirit, the second moment of $T_0$ can be derived for $\nu>2$ by setting $n=2$ in \eref{eq:MFPTn_r0} as $\left<T_0^2\right>=x_0^2/[D_0^2(\nu-1)(\nu-2)]$.
The standard deviation in $T_0$, $\sigma(T_0)\coloneqq \sqrt{\left<T_0^2\right>-\left<T_0\right>^2}$ thus reads 
\begin{equation}
\sigma(T_0)=
\frac{x_0}{D_0(\nu-1)}\sqrt{ \frac{1}{\nu-2}}.
\label{eq:STDV0}
\end{equation} 
\indent
Recalling that $\mu=D_0(\nu-1)$, we see that when the potential is repulsive ($\mu< 0$) or when the particle freely diffuses ($\mu=0$) with a diffusion coefficient $D(x)=D_0 |x|$, it takes infinite mean time to reach the origin in the limit $r\to0$. \eref{eq:MFPT0_r0}, on the other hand, shows that the particle can reach the origin in finite mean time only when the potential is attractive ($\mu>0$), which analytically validates our physical intuition. \\
\indent
Comparing Eqs. (\ref{eq:MFPT0_r0}) and (\ref{eq:STDV0}), we see that for $1<\nu\le2$, the mean is finite  whereas the standard deviation diverges. Eqs. (\ref{eq:MFPT0_r0}) and (\ref{eq:STDV0}) also indicate that $\sigma(T_0)=\left<T_{0}\right>$ when $\nu=3$. For $2<\nu<3$, $\sigma(T_0)$ and $\left<T_{0}\right>$ are both finite, but $\sigma(T_0)>\left<T_{0}\right>$, which means that the fluctuations in the first-passage time $T_0$ around its mean are high. In contrast, for $\nu>3$, $\sigma(T_0)<\left<T_{0}\right>$, i.e, the fluctuations in $T_0$ around $\left<T_{0}\right>$ are less. Therefore, the persistent exponent $\nu\coloneqq(1+\mu D_0^{-1})$, which characterizes the nature (attractive or repulsive) and relative strength of the potential (manifested by the drift $\mu$) over diffusion (manifested by $D_0$), marks the signature of different dynamical behaviors for the underlying process. Next, we investigate how the introduction of stochastic resetting affect the dynamics.

\subsection{First-passage with resetting}
When the absorbing boundary is placed at the origin, the survival probability in presence of resetting can be obtained from \eref{eq:Qr_1} 
utilizing the liming expression $\lim_{x_a\to 0}x_a^{\nu/2}K_{\nu}(\alpha\sqrt{x_a})\simeq 2^{\nu-1}\Gamma(\nu)/\alpha^{\nu}$ as before, and that gives
\begin{align}
\tilde{Q_r}(s|x_0)=\frac{1-\frac{2x_0^{\frac{\nu}{2}}
}{\Gamma(\nu)}\left[\frac{s+r }{D_0}\right]^{\frac{\nu}{2}}K_{\nu}\left(2\sqrt{\frac{s+r}{D_0}}\sqrt{x_0}\right)}{s+r\left[\frac{2x_0^{\frac{\nu}{2}}
}{\Gamma(\nu)}\left[\frac{s+r }{D_0}\right]^{\frac{\nu}{2}}K_{\nu}\left(2\sqrt{\frac{s+r}{D_0}}\sqrt{x_0}\right)\right]}.
\label{eq:Qr0}
\end{align}
The associated mean FPT can either be obtained directly from \eref{eq:MFPT_1} utilizing the above liming expression for $x_a\to 0$ or setting $s=0$ in \eref{eq:Qr0}, and it reads
\begin{align}
\left<T_r\right>=\frac{1}{r}\left[\frac{\Gamma(\nu)}{2 \left[\frac{r x_0}{D_0}\right] ^{\frac{\nu}{2}} K_{\nu}\left(2\sqrt{\frac{r x_0}{D_0}}\right)}-1\right].
\label{eq:MFPT0}
\end{align}
Recalling the definition of the Gamma function, we observe from \eref{eq:MFPT0} that $\left<T_r\right>$ is finite for $\nu>0$. Comparing with \eref{eq:MFPT0_r0}, we see that while the mean FPT to the origin diverges for $0<\nu\le1$ (i.e, when the potential is either nonexistent or weakly repulsive) for the underlying process, it becomes finite in presence of resetting.\\
\indent
In a similar spirit as in the case of  $\left<T_r\right>$, the standard deviation of the FPT can be calculated directly from \eref{eq:STDV_1} using the limiting expression for $x_a\to0$. Alternatively, it can be derived by first calculating the second moment from \eref{eq:Qr0} and then utilizing \eref{eq:MFPT0}. In either way it leads to
\begin{align}
\sigma(T_r)=\frac{1}{r}\sqrt{
\frac{\Gamma(\nu)\left[\Gamma(\nu)-
4\left[\frac{r x_0}{D_0}\right]^{\frac{1+\nu}{2}}
K_{\nu-1}\left(2 \sqrt{\frac{r x_0}{D_0}}
\right)\right]}
{4\left[\frac{r x_0}{D_0}\right]^{\nu}\left[K_{\nu}\left(2 \sqrt{\frac{r x_0}{D_0}}\right)\right]^2}-1}.
\label{eq:STDVr0}
\end{align}
\indent
In \fref{Fig3}, we plot the mean FPT and the standard deviation of the FPT as functions of $r$ from Eqs.~(\ref{eq:MFPT0}) and (\ref{eq:STDVr0}), respectively, for different values of the persistent exponent, $\nu$.
It shows that for $r\to0$, the mean $\left<T_r\right>$ and the standard deviation $\sigma(T_r)$ of the FPT, both diverge for $0<\nu<1$ [panel (a)], while for $1<\nu<2$ the mean FPT is finite, but the standard deviation diverges [panel (b)]. For $\nu>2$, both the mean and the standard deviation become finite for $r\to 0$, though $\sigma(T_{r\to0})>\left<T_{r\to0}\right>$ for $2<\nu<3$ [panel (c)], whereas $\sigma(T_{r\to0})<\left<T_{r\to0}\right>$ for $\nu>3$ [panel (d)]. Therefore, all the results in the limit $r\to 0$ are in agreement with our previous derivations.\\
\indent
Panels (a)$-$(c) of \fref{Fig3} suggest that when $0<\nu<3$, $\left<T_r\right>$ shows a non-monotonic variation with the resetting rate. This implies that when the potential is either weakly repulsive or weak to moderately attractive, resetting expedites the first-passage of the particle to the origin. In contrast, panel (d) of \fref{Fig3} shows that for $\nu>3$, the mean FPT monotonically increases with $r$, thereby suggesting that resetting can only delay the first-passage to the origin when the potential is strongly attractive. This marks the signature of a resetting transition\cite{Restart-Biophysics1,exponent,RayReuveniJPhysA, Landau}, which is expected as $\nu$ increases beyond a tipping point. In the following Section, we present a comprehensive analysis of the resetting transition for the present system. 
\section{The resetting transition}
We observe from \fref{Fig3} that as $\nu$ increases, the variation of the mean FPT with the resetting rate changes from non-monotonic to monotonic. In other words, the optimal resetting rate, i.e., the rate of resetting that minimizes the mean FPT, is non-zero for smaller values of $\nu$ and it reduces to zero as $\nu$ grows. This indicates that the optimal resetting rate, denoted $r^{\star}$, should serve as a suitable observable to study the resetting transition. Motivated by this idea, we proceed to explore the resetting transition in terms of the optimal resetting rate.
\subsection{The optimal resetting rate}
\begin{figure*}[t!]
\begin{centering}
\includegraphics[width=5.9cm]{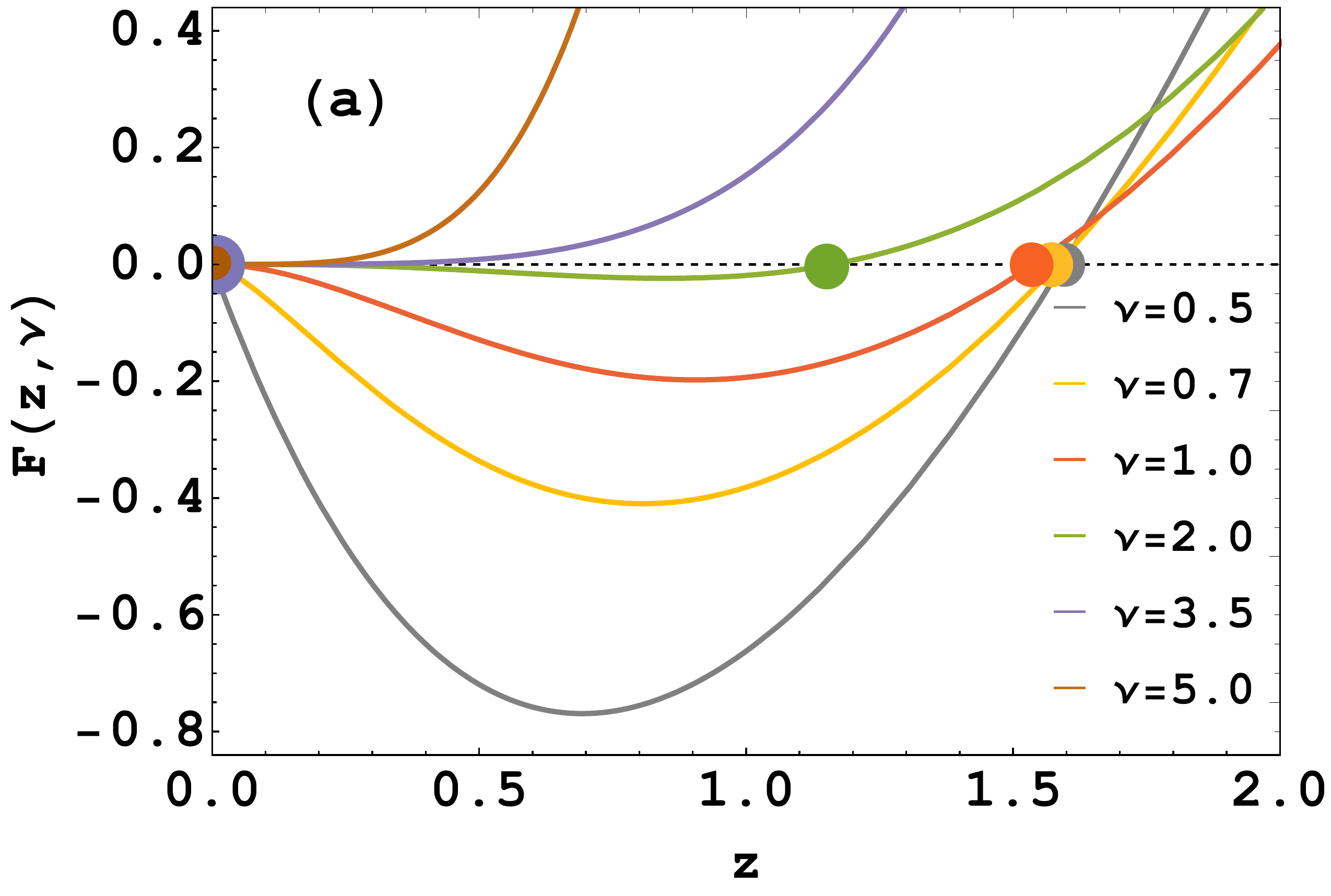}
\includegraphics[width=5.9cm]{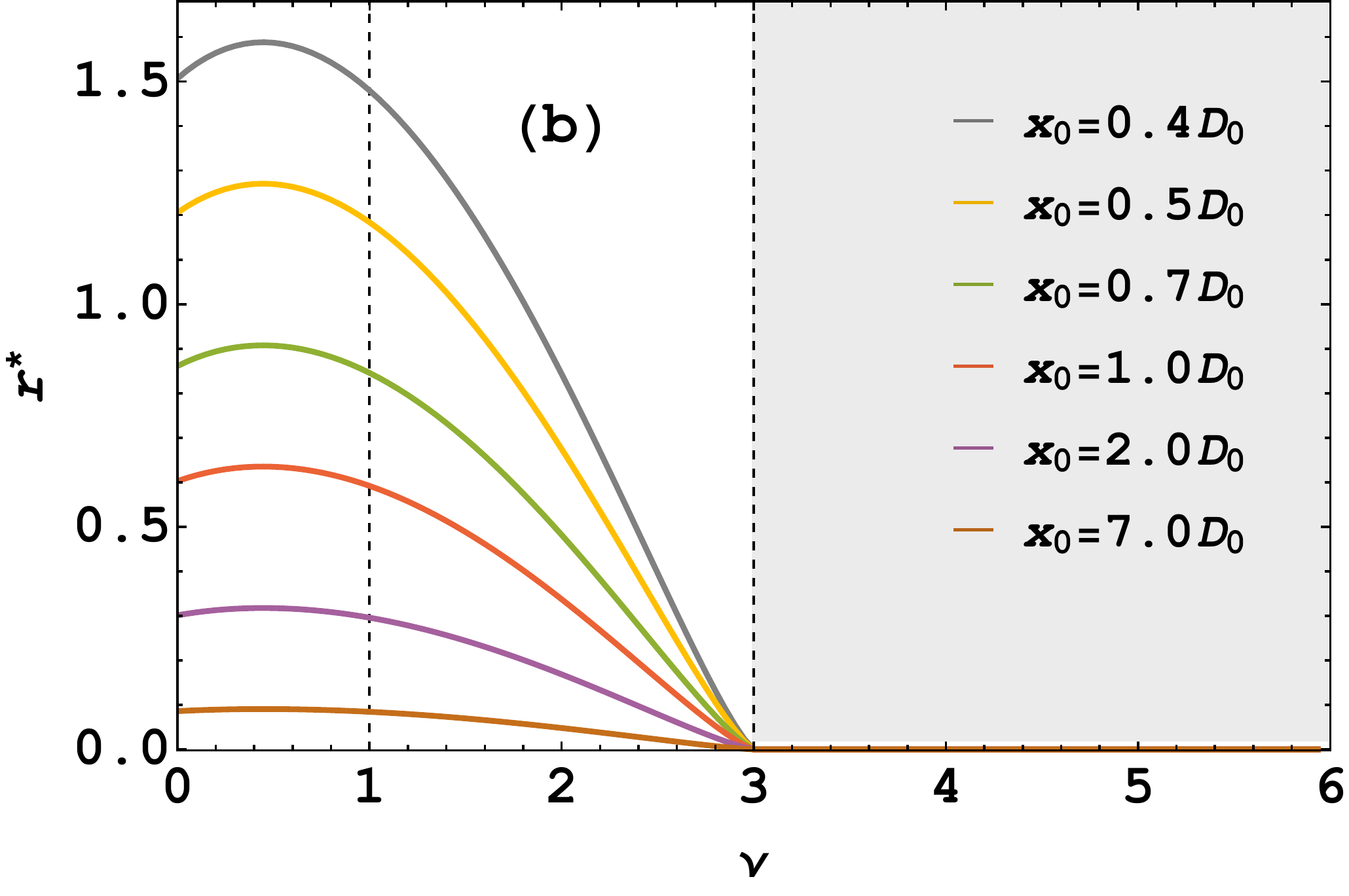}
\includegraphics[width=5.85cm]{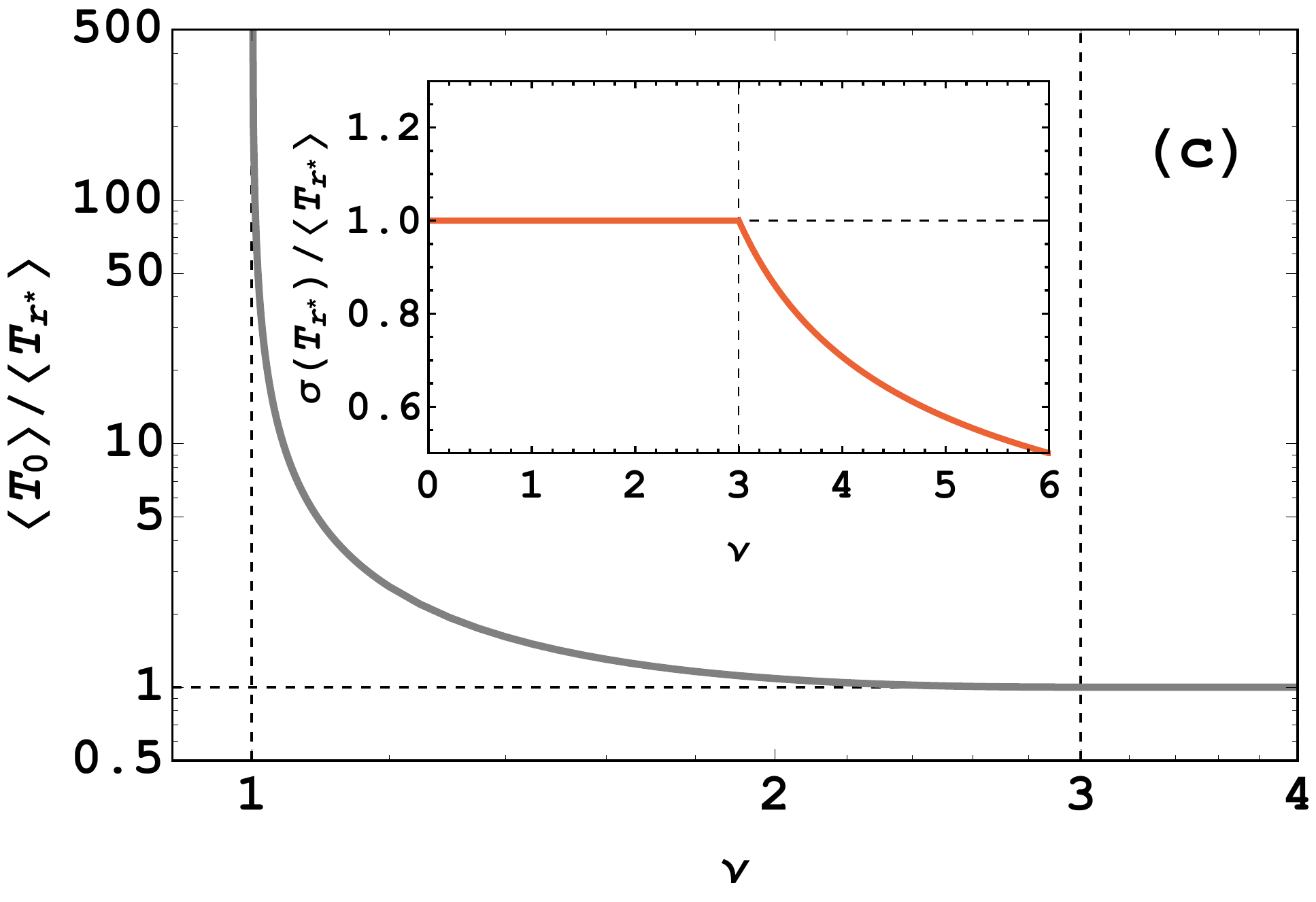}
\end{centering}
\caption{Panel (a): Graphical solution of \eref{eq:trans} for different values of $\nu$. The solutions, $z^{\star}$, are marked by colored circles.
Panel (b): The optimal resetting rate $r^{\star}$ vs. $\nu$ for different values of $x_0 D_0^{-1}$, calculated by plugging in $z^{\star}$ into \eref{eq:r_z}. For $\nu<1$, the potential is repulsive, whereas it is attractive for $\nu>1$. The non-zero $r^{\star}$ values for $\nu<3$ (white regime) indicate that here resetting expedites the first-passage to the origin. In contrast, $r^{\star}=0$ values for $\nu\ge3$ implies that resetting can no longer accelerate the first-passage process (gray regime). This leads to the resetting transition at $\nu=3$.  
Panel (c): Main: The maximal speedup $\left<T_0\right>/\left<T_{r^{\star}}\right>$ vs. $\nu$ from \eref{eq:maxsp}. For $\nu\le1$, the maximal speedup is infinite, and it decays to unity at $\nu =3$, the point of resetting transition. Inset: The relative fluctuations in FPT for optimal resetting, $\sigma(T_{r^{\star}})/\left<T_{r^{\star}}\right>$, vs. $\nu$ from \eref{eq:CVstar}, indicating that the resetting transition occurs at $\nu=3$.
}
\label{Fig4}
\end{figure*}

In order to study the optimal resetting rate, we define a new variable
$z\coloneqq \alpha \sqrt{x_0}\equiv2\sqrt{rx_0/D_0}$, which leads to
\begin{eqnarray}
r=\frac{z^2 D_0}{4 x_0}.
\label{eq:r_z}
\end{eqnarray}
\eref{eq:MFPT0} in terms of $z$ reads
\begin{align}
\left<T_r\right>=\left(\frac{4 x_0}{D_0 z^2}\right)\left[\frac{2^{\nu-1}\Gamma(\nu)}{z^{\nu}K_{\nu}(z)}-1\right].
\label{eq:MFPT_z}
\end{align}
Since $r^{\star}$ minimizes the mean FPT, we have $[d \left<T_r\right>/dr]_{r=r^{\star}} = 0$. \eref{eq:r_z} suggests that $d\left<T_r\right>/dr=
(2x_0D_0^{-1}/z)[d \left<T_r\right>
/dz]$. Differentiating \eref{eq:MFPT_z} with respect to $z$ we obtain the following transcendental equation 
\begin{align}
F(z,\nu)\coloneqq4z^{\nu}K_{\nu}(z)+2^{\nu}\Gamma(\nu)\left[\frac{z K_{\nu-1}(z)}{ K_{\nu}(z)}-2\right]=0.
\label{eq:trans}
\end{align}
\indent
In \fref{Fig4}(a) we graphically solve \eref{eq:trans}. The solutions, $z^{\star}$, when plugged in into \eref{eq:r_z}, give the optimal resetting rate $r^{\star}$. In \fref{Fig4}(b) we plot $r^{\star}$ vs. $\nu$ for different values of $x_0 D_0^{-1}$ to find that $r^{\star}$ is non-zero only when $\nu<3$. This implies that resetting expedites the first-passage to the origin for weakly repulsive/attractive potential. In contrast, $r^{\star}$ becomes zero for $\nu\ge 3$, which means that resetting can no longer assist the first-passage to the origin when the potential becomes strongly attractive. The optimal resetting rate thus marks the point of resetting transition at $\nu=3$ (i.e., $\mu=2D_0$). Recalling Eqs. (\ref{eq:MFPT0_r0}) and (\ref{eq:STDV0}), we see that resetting accelerates the first-passage of the particle to the origin when the fluctuations in the FPT of the underlying process around its mean are high. This agrees with the general theory of first-passage with resetting \cite{Restart-Biophysics1,ReuveniPRL}. \fref{Fig4}(b) clearly shows that the condition of resetting transition is unaffected by the distance of the initial position $x_0$ from the origin. This is strikingly different from homogeneous diffusion in a linear potential, where the condition reads\cite{RayReuveniJPhysA} $\mu=2D/x_0$, $D$ being the constant diffusion coefficient.  \\
\indent
\fref{Fig4}(b) also exhibits an interesting non-monotonic behavior of $r^{\star}$ for $\nu<1$, where the linear potential is repulsive in nature. Starting from $\nu\to 0^+$, $r^{\star}$ increases to attain a maximum, and then gradually decreases until it vanishes at $\nu=3$. This initial rise is apparently counter-intuitive, but it can be physically understood as follows. In the present context, the particle undergoes {\it space-dependent} diffusion; hence the fluctuations in its movement get much more prominent as its distance from the origin increases. It moves  less erratically as it approaches $x=0$, which makes that regime drift-dominated. As discussed earlier, resetting can expedite the first-passage to the origin by cutting short the trajectories that tend to move away from $x=0$. 
In addition, it might rescue the particle, somewhat trapped in the interval $0<x<x_0$ because of the low effective diffusion, especially when the drift velocity is weak. The latter role of resetting is somewhat analogous to the action of a repulsive potential ($\mu<0$), as it drives the particle away from the origin. This explains the initial increase of $r^{\star}$ as the repulsion weakens, i.e., $\nu$ grows, until a tipping point. After that, the role of resetting in minimizing the lifetime of the trajectories where the particle diffuses away from the origin ($x\gg x_0$) becomes predominant and the optimal resetting rate gradually decreases with $\nu$. The non-monotonic variation of $r^{\star}$ for smaller values of $\nu$ can thus be attributed to the space-dependent nature of the diffusion. Next, to complete the discussion on the optimal resetting rate, we explore how $r^{\star}$ decays near the point of resetting transition. \\
\indent
In the recent works on  the general theory of first-passage with resetting \cite{ReuveniPRL,PalReuveniPRL,exponent,Landau}, it has been shown that the mean FPT for a process with resetting can be written as a polynomial of the resetting rate $r$ as
\begin{align}
\left<T_r\right>=a_0+a_1r+a_2r^2+O(r^3)+\cdots, 
\label{eq:mfpt_FPUR2}
\end{align}
where the coefficients of such expansion are functions of the moments of the FPT distribution of the underlying process. In particular, we have\cite{exponent,Landau} $a_0=\left<T_0\right>$, $a_1=\left<T_0\right>^2-\frac{1}{2}\left<T_0^2\right>$ and $a_2=\frac{1}{6}\left<T_0^3\right>+\left<T_0\right>^3-\left<T_0\right>\left<T_0^2\right>$. For the optimal resetting rate $r^{\star}$ we have  $[d\left<T_r\right>/dr]_{r=r^{\star}}=0$, which gives \cite{exponent,Landau}
\begin{align}
r^{\star}\simeq \frac{|a_1|}{2a_2}=\frac{\left<T_0\right>^2-\frac{1}{2}\left<T_0^2\right>}{\frac{1}{6}\left<T_0^3\right>+\left<T_0\right>^3-\left<T_0\right>\left<T_0^2\right>}.
\label{eq:rstar_T0}
\end{align}
Utilizing \eref{eq:MFPTn_r0} to evaluate the moments of $T_0$ and recalling that $\nu\coloneqq(1+\mu D_0^{-1})$, from \eref{eq:rstar_T0} we obtain
\begin{align}
r^{\star}=\frac{\mu}{4x_0}\left[\frac{(\mu D_0^{-1}-2)^2}{\left[\frac{1}{6}\mu D_0^{-1}-1\right]\mu D_0^{-1}+2}\right],
\label{eq:rstar_SDDR}
\end{align}
which holds for smaller values of $r$. Expansion of the right hand side of \eref{eq:rstar_SDDR} in terms of $\eta\coloneqq \mu D_0^{-1}$ leads to $r^{\star}=\frac{D_0}{2x_0}\eta-\frac{ D_0}{4x_0}\eta^2-\frac{ D_0}{24x_0}\eta^3+\cdots$, i.e, $r^{\star}\simeq \frac{\mu}{4x_0}(2-\mu D_0^{-1})$. This allows us to write
\begin{align}
r^{\star}\propto (\nu_{c}-\nu)^{\beta},
\label{eq:rstar_exponent}
\end{align}
where $\nu_{c}=3$ is the point of resetting transition and $\beta=1$ is the critical exponent. \eref{eq:rstar_exponent} shows that the resetting transition observed here is continuous and $r^{\star}$ decays with an exponent equal to unity near the point of transition.\\
\indent

\subsection{The maximal speedup}
The optimal resetting rate minimizes the mean FPT and thereby leads to the maximal speedup for the resulting first-passage process. This inspires us to quantify the maximal speedup as the ratio between the mean FPT for the underlying process and the process under optimal resetting. 
Utilizing \eref{eq:MFPT0_r0} and \eref{eq:MFPT_z}, we can write
\begin{align}
\frac{\left<T_{0}\right>}{\left<T_{r^{\star}}
\right>}=
\begin{cases}
\frac{{z^{\star}}^{2+\nu}K_{\nu}(z^{\star})}{4\left[2^{\nu-1}\Gamma(\nu)-{z^{\star}}^{\nu}K_{\nu}(z^{\star})\right]}\;\;\;\mbox{for}\;\;\nu<3,\\ 
1\;\;\;\mbox{\hspace{2.6cm}for}\;\;\nu\ge3,
\end{cases}
\label{eq:maxsp}
\end{align}
where we considered the fact that for $\nu\le3$, $z^{\star}=0$ and plugged in that into \eref{eq:MFPT_z}.\\
\indent
In \fref{Fig4}(c), we plot the maximal speedup from \eref{eq:maxsp} vs. $\nu$, which shows that the introduction of resetting renders the infinite mean FPT of the underlying process finite for $\nu\le1$, leading to infinite speedup. For $1<\nu<3$, the maximal speedup is finite but greater than unity, which suggests that resetting still expedites the first-passage in this case. In contrast, for $\nu\le3$ it becomes unity that implies that the underlying process ends faster compared to that with resetting in this regime. Next, to complete the analysis of the resetting transition, we investigate the stochastic fluctuations in the first-passage time at the optimal resetting rate.
\subsection{The fluctuations in FPT for optimal resetting}
Recalling that $z\coloneqq\sqrt{rx_0/D_0}$, and plugging in the same into \eref{eq:STDVr0}, we obtain an expression of $\sigma(T_r)$ as a function of $z$ as
\begin{align}
\sigma(T_r)=\left(\frac{4 x_0}{D_0 z^2}\right)\sqrt{
\frac{\Gamma(\nu)\left[\Gamma(\nu)-
4\left[\frac{z}{2}\right]^{1+\nu}
K_{\nu-1}\left(z\right)\right]}
{4\left[\frac{z}{2}\right]^{2\nu}\left[K_{\nu}\left(z\right)\right]^2}-1}.
\label{eq:STDVz}
\end{align}
Combining \eref{eq:MFPT_z} with \eref{eq:STDVz}, we can readily express the relative stochastic fluctuations, defined as $\sigma(T_{r})/\left<T_{r}\right>$, in terms of $z$. 
Setting $z=z^{\star}$ [solutions of \eref{eq:trans}] we obtain the identity  $\left(z^{\star}/2\right)^{1+\nu}\Gamma(\nu)K_{\nu-1}(z^{\star})\equiv\left(z^{\star}/2\right)^{\nu}\Gamma(\nu)K_{\nu}(z^{\star})-2\left(z^{\star}/2\right)^{2\nu}[K_{\nu}(z^{\star})]^2$, and incorporating that in the expression of $\sigma(T_{r})/\left<T_{r}\right>$, we finally get
\begin{align}
\frac{\sigma(T_{r^{\star}})}{\left<T_{r^{\star}}
\right>}=
\begin{cases}
1\;\;\;\mbox{\hspace{1cm}for}\;\;\nu<3\\
\frac{1}{\sqrt{\nu-2}}\;\;\;\mbox{\hspace{0.4cm}for}\;\;\nu\ge3.\\
\end{cases}
\label{eq:CVstar}
\end{align}
Here we utilized the results for the underlying process from Eqs.~(\ref{eq:MFPT0_r0}) and (\ref{eq:STDV0}) for $\nu\ge3$. In the inset of \fref{Fig4}(c), we plot the relative fluctuations to observe the signature of the resetting transition in terms of it.\\
\indent
\eref{eq:CVstar} proves analytically that for $\nu<3$, the stochastic fluctuations around the mean FPT is always unity for optimal resetting. This agrees with the results established by the general theory of first-passage with resetting\cite{ReuveniPRL}. Looking back at panels (a)$-$(c) of \fref{Fig3}, we see that the curves representing the mean FPT and the standard deviation of the FPT always intersect at the optimal resetting rate, which supports \eref{eq:CVstar}. \vspace{-0.4cm}
\begin{figure*}[ht!]
\begin{centering}
\includegraphics[width=18.0cm]{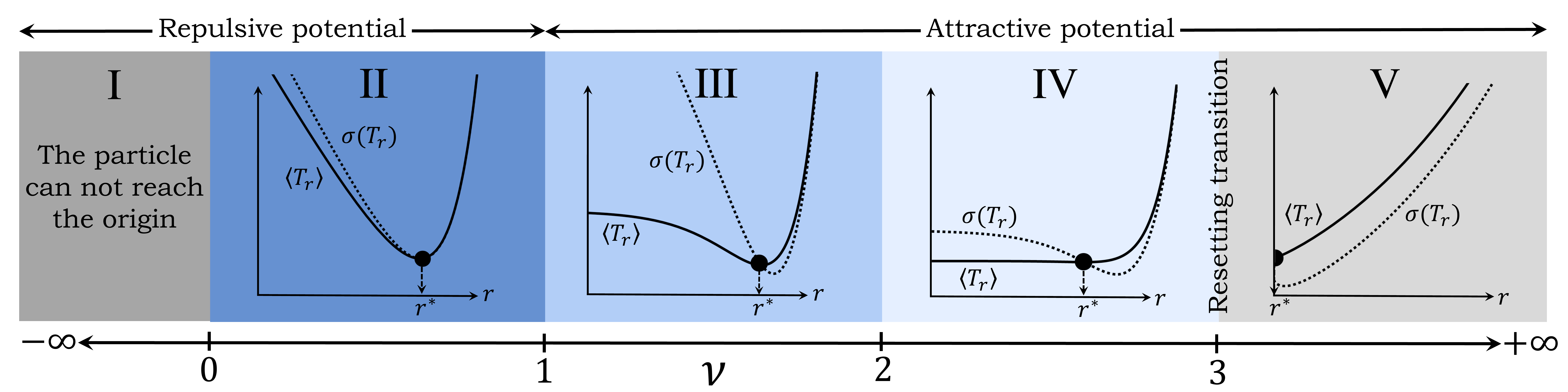}
\end{centering}
\caption{A phase diagram showing the possible effects of stochastic resetting on the first-passage of a particle to the origin, when it undergoes space-dependent diffusion with diffusion coefficient $D(x)=D_0|x|$ in presence of a constant bias $\mu$. Transitions between the different phases occur by tuning the parameter $\nu\coloneqq(1+\mu D_0^{-1})$, which captures the interplay between the drift and the diffusion. Phase I: Here the particle never reaches the origin. Phase II-V: The particle eventually reaches the origin. However, the behavior of the mean  $\left<T_{r}\right>$ and the standard deviation $\sigma(T_{r})$ of the FPT are markedly different for the different phases in the absence of resetting ($r\to0$), as discussed in details in the main text. In phases II–IV, the introduction of
resetting reduces the mean FPT, i.e., expedites the first-passage of the particle to the origin, whereas in phase V it is the other way around. The transition between these
two opposite behaviors occurs at $\nu = 3$.}
\label{Fig5}
\end{figure*}
\section{The phase diagram}
In the previous Sections, we have thoroughly explored the possible ways with which stochastic resetting can affect the first-passage of a particle to the origin, when the particle undergoes space-dependent diffusion in presence of a constant bias. In doing so, we observed that such effect of resetting is guided solely by the parameter $\nu\coloneqq(1+\mu D_0^{-1})$, the persistent exponent of the underlying process. This allows us to construct a full phase diagram for the present problem, as displayed in \fref{Fig5}. \\
\indent
From \fref{Fig5}, we observe that the entire range of $\nu\in\{-\infty,\infty\}$ can be divided into five distinct phases and dynamical transitions occur between these phases when $\nu$ is tuned. For $-\infty<\nu\le0$ the potential is strongly repulsive, and the particle never reaches the origin in that case, which leads to phase I in \fref{Fig5}. \\
\indent
For weakly repulsive potential ($0<\nu<1$) and in absence of any bias ($\nu=1$), resetting renders the infinite mean FPT for the underlying process finite and this marks phase II. When the potential is weakly attractive ($1<\nu\le2$), the mean FPT of the underlying process is finite, but introduction of resetting decreases it further, which constructs phase III. Note that both in phase II and III, the infinite standard deviation of the FPT for the underlying process becomes finite due to resetting. \\
\indent
For weak to moderately attractive potential ($2<\nu\le3$), for $r\to0$ both the mean FPT and the standard deviation of the FPT are finite, but $\sigma(T_0)>\left<T_0\right>$. Resetting still expedites the first-passage to the origin in this regime, which is displayed as phase IV in \fref{Fig5}. Summarizing the above results, we see that resetting accelerates the first-passage of the particle to the origin for phases II$-$IV [marked in different shades of blue in \fref{Fig5}], which is confirmed by the non-monotonic variation of $\left<T_r\right>$ with the resetting rate and the non-zero optimal resetting rates marked by the point of intersection of the curves representing $\left<T_r\right>$ and $\sigma(T_r)$ in each phase.\\
\indent
In contrast, for $3<\nu<\infty$, i.e, when the potential is strongly attractive, in absence of resetting $\sigma(T_0)<\left<T_0\right>$. Introduction of resetting delays the first-passage to origin in this regime, which marks phase V of \fref{Fig5}. This phase can be identified from the monotonic increase in $\left<T_r\right>$ with $r$ and the fact that the curves representing $\left<T_r\right>$ and $\sigma(T_r)$ do not intersect here (the optimal resetting rate is zero, as marked in \fref{Fig5}). The resetting transition thus occurs at $\nu=3$, where the mean FPT of the underlying process is exactly equal to the standard deviation of the FPT.
\section{Conclusions}
In this article, we presented a comprehensive analysis of the effect of stochastic resetting on the first-passage properties of heterogeneous diffusion in presence of a constant bias. In our model, a particle that diffuses in a potential $U(x)\propto \mu|x|$ with a space-dependent diffusion coefficient $D(x)= D_0 |x|$, is subject to stochastic resetting with a constant rate $r$. Assuming an absorbing boundary placed at a position $x_a<x_0$, where $x_0>0$ is the initial position of the particle, we derived an exact expression of the survival probability in the Laplace space and subsequently explored its first-passage to the origin as a limiting case of that general result. \\
\indent
In that limit, i.e., when $x_a\to 0$, we first presented an in depth analysis of the underlying process ($r\to 0$) that includes the derivation of an exact analytic expression of the first-passage time distribution. When subjected to resetting, the system is observed to undergo a series of dynamical transitions as a single parameter $\nu\coloneqq(1+\mu D_0^{-1})$, is tuned. For $\nu<0$, when the potential is strongly repulsive, the particle never reaches the origin. Resetting is expected to generate a non-equilibrium steady state in that case, which we plan to study elsewhere. For $\nu>0$ the potential is either weakly repulsive or attractive, and the particle can eventually reach the origin. Resetting accelerates the completion of the associated first-passage process for $\nu<3$, but delays it when $\nu\ge3$. We provided a detailed account of the resetting transition observed at $\nu=3$. \\
\indent
Space-dependent diffusion naturally gives rise to a noise that is multiplicative in nature, and is crucial in modeling a large variety of diffusion processes. We are hopeful that the present study will inspire a series of theoretical and experimental works that bring together heterogeneous diffusion and stochastic resetting.
\vspace{-0.2cm}
\section*{ACKNOWLEDGEMENTS}
\noindent
The author acknowledges support from the Raymond and Beverly Sackler Center for Computational Molecular and Materials Science, Tel Aviv University.
\vspace{-0.2cm}
\section*{DATA AVAILABLITY}
\noindent
The data that supports the findings of this work are available within the article.

\end{document}